# Correlating Chemical Reaction and Mass Transport in Hydrogen-based Direct Reduction of Iron Oxide

Xueli Zheng[1,5], Subhechchha Paul[1], Lauren Moghimi[1], Yifan Wang[1], Rafael A. Vilá[1], Fan Zhang[2], Xin Gao[1], Junjing Deng[3], Yi Jiang[3], Xin Xiao[1], Chaolumen Wu[4], Louisa C. Greenburg[1], Yufei Yang[1], Yi Cui[1], Arturas Vailionis[1], Ivan Kuzmenko[3], Jan llavsky[3], Yadong Yin[4], Yi Cui[1,5], Leora Dresselhaus-Marais[1,6]*

[1]Department of Materials Science and Engineering, Stanford University, Stanford, California 94305, USA.

[2]Materials Measurement Science Division, National Institute of Standards and Technology, Gaithersburg, MD 20899, USA.

[3]X-ray Science Division, Advanced Photon Source, Argonne National Laboratory, Argonne, Illinois 60439, USA.

[4]Department of Chemistry, University of California, Riverside, CA 92521, USA.

[5]Stanford Institute for Materials and Energy Sciences, SLAC National Accelerator Laboratory, Menlo Park, California, USA.

[6]Department of Photon Science, SLAC National Accelerator Laboratory, 2575 Sand Hill Rd, Menlo Park, 94025, CA, USA.

*Correspondence and requests for materials should be addressed to L.D.M (leoradm@stanford.edu)

**Steelmaking contributes 8% to the total $CO_2$ emissions globally, primarily due to coal-based iron ore reduction. Clean *hydrogen*-based ironmaking has variable performance because the dominant gas-solid reduction mechanism is set by the defects and pores inside the mm-nm sized oxide particles that change significantly as the reaction progresses. While these governing dynamics are essential to establish continuous flow of iron and its ores through reactors, the direct link between agglomeration and chemistry is still contested due to missing measurements. In this work, we directly measure the connection between chemistry and agglomeration in the smallest iron oxides relevant to magnetite ores. Using synthesized spherical 10-nm magnetite particles reacting in $H_2$, we resolve the formation and consumption of wüstite (FeO) - the step most commonly attributed to agglomeration. Using X-ray scattering and microscopy, we resolve crystallographic anisotropy in the rate of the initial reaction, which becomes isotropic as the material sinters. Complementing with imaging, we demonstrate how the particles self-assemble, subsequently react and sinter into ~100x oblong grains. Our insights into how morphologically uniform iron oxide particles react and agglomerate $H_2$ reduction enable future size-dependent models to effectively describe the multiscale iron ore reduction.**

**Introduction**

Steel is ubiquitous in modern society, but its production contributes to 8% of global $CO_2$ emissions (*1-3*). More than half of the $CO_2$ emitted in steelmaking occurs when reducing the iron oxide ores into molten iron. Ironmaking is now done predominantly using blast furnaces that use processed coal to form the carbonaceous reducing agents, generating vast $CO_2$ emissions (*4, 5*). Gas injection has been exploited to reduce coal requirements and $CO_2$ emissions, but coal is hard to be fully removed from blast-furnace ironmaking. Efforts have emerged to use natural gas (*6, 7*), electrochemistry (*8*), and hydrogen gas (*9, 10*) for cleaner ironmaking, but those only make up ~5% of the global market today (*11*).

Hydrogen based direct reduction (HyDR) of iron ores is a method of clean ironmaking that replaces coal with $H_2$ gas, generating $H_2O$ instead of $CO_2$. While seemingly simple, the HyDR process has been slow to commercialize because of its endothermic chemistry, which requires the reaction to be operated at higher temperatures (above 1000°C) (*12*). However, high temperature causes the iron ores and product iron to agglomerate and clog the reactors due to a process called sintering (*13-15*). Models of the sintering process are known to be essential to effectively scale HyDR, but the high variability between different ores and local reactor conditions has yielded challenges in accuracy (*9, 16*).

HyDR reactors typically operate at 800-1000 °C, but the endothermic chemistry causes microscopic temperature fluctuations inside the macroscopic vessels that change the local structure (*15*). Whiskering is known to proceed by first overcoming a relatively high activation energy to nucleate, though it proceeds via a lower-energy growth mechanism (*17*). As such the whiskering process is highly dependent on small distortions that seed further accumulation. As such, understanding the anomalous low-temperature mechanisms through which kinetics may link to whiskering is essential to design efficient reactors.

It has been shown that accurate models must describe both the microstructure and chemistry of the reduction over time (*3, 18*). Industrial ironmaking progresses as hematite ($Fe_2O_3$) to magnetite ($Fe_3O_4$) to wüstite (FeO) to iron (Fe), with the rate-limiting step being the FeO to Fe transition (*19, 20*). The proposed surface-based (topochemical) mechanism of wustite to iron conversion is the depletion of sub-stoichiometric oxygen (via diffusion) until it reaches a critical threshold to nucleate regions of the Fe phase (*21*). Supporting the topochemical models, several groups have observed kinetics differences at surface-based defects, i.e. boundaries between phases and grains (*15, 22*). By contrast, early experiments demonstrated how the reduction rate of iron oxide ores also depends strongly on the evolution of internal pores even in initially dense samples (*23*). Significant volume reduction on each step of the reactions generates new defects that are "inherited" by the reduced oxides and alter the rates of subsequent reduction steps. To date, most have studied these relations in the mm-scale "pellets" used in most of today's reactor designs. The refined iron ores, or "fines," consolidate the process, but their 100-um to 10-nm sized particles are typically overlooked.

In this work, we present a multiscale view of HyDR chemistry using nanoparticles with advanced microscopy and scattering methods to directly map the formation of wüstite and then iron and its link to sintering. We study pure iron oxide particles at nanometer scales representative of the smallest particles in industrial iron ore fines. For comparison, we used controlled lab-synthesized 10-nm particles and compared them to industrial magnetite ore fines with $H_2$ reduction. Using *in-situ* X-ray diffraction (XRD), we demonstrate the kinetics of the two steps of reduction ($Fe_3O_4 \rightarrow FeO$, then $FeO \rightarrow Fe$) and observe unexpected FeO phase formation at 300 °C while it is not thermodynamically stable. A crystallographic look at the chemistry resolves that the formation and consumption of wüstite occur with rates that differ depending on the energy of each crystallographic surface (facet) at 300 °C. Simultaneous small-angle X-ray scattering (SAXS) indicated that as the reduction occurs, the particle sizes dramatic increase, indicating the onset of the sintering process. Images of our samples by transmission electron microscopy (TEM) and X-ray ptychographic tomography at the nano/microscopic scales reveal a complex hierarchical organizational process that precedes the chemistry and sintering steps. Our study of pristine iron oxide nanoparticles affords a unique view of how chemistry links to multiscale mechanics in the absence of any impurities, initial porosity or shape effects. This insight provides a key starting point to connect length scales in models to design and optimize different HyDR approaches.

**Results**

**Comparison of industrial and nanoparticle magnetite reduction.** It is challenging to connect the chemistry of the reaction from the nanoscale to microscale due to complexity of the reaction and defects in natural samples. Therefore, we begin by comparing magnetite concentrate sample extracted from the Mesabi Range in Minnesota and lab-synthesized $Fe_3O_4$ nanoparticles to confirm that the nanoscale chemistry is representative. The scanning electron microscopy images of industrial iron ore fines ($Fe_3O_4$) shown in **Figure 1a** display the wide range of particle size distribution from nanoscale to microscale (additional images shown in **Figure S1**). In comparison, we synthesized uniform spherical $Fe_3O_4$ nanoparticles with a 10-nm diameter (**Figure 1d**). For the solid-gas reaction, XRD measures the phase transformations based on the amount of each crystal structure observed, as shown in **Figure 1**. We performed *in-situ* XRD for these reductions in an isothermal reactor at T = 800 °C using 2% $H_2$/Ar for industrial (**Figure 1b**) and nanoparticle (**Figure 1e**) $Fe_3O_4$. *In-situ* XRD showed the reduction of magnetite ($Fe_3O_4$) to wüstite (FeO), and subsequently FeO to Fe. We observed the phase fractions of $Fe_3O_4$, FeO, and Fe for industrial $Fe_3O_4$ (**Figure 1c**) and nanoparticle $Fe_3O_4$ (**Figure 1f**). Mapping the X-ray diffraction signal, we observed a rapid reduction of $Fe_3O_4$ to FeO for both industrial $Fe_3O_4$ (**Figure 1c**) and nanoscale $Fe_3O_4$ (**Figure 1f**). Phase fraction results revealed the fast reduction from $Fe_3O_4$ to FeO. The rate-limiting step is the reduction of FeO to metallic Fe for both industrial $Fe_3O_4$ and nanoscale $Fe_3O_4$. Our industrial $Fe_3O_4$ fines exhibited an initial lag time for the first 90 minutes at 800 °C (**Figure 1e**), indicating slow hydrogen diffusion rates in the industrial $Fe_3O_4$.

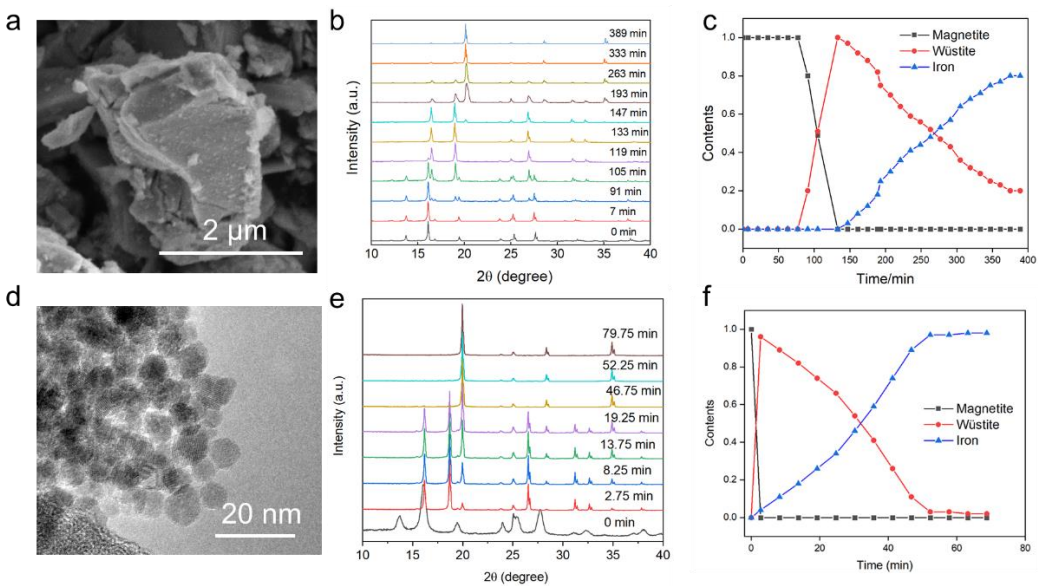

**Figure 1.** Morphology and reaction kinetics of industrial $Fe_3O_4$ and nanoscale $Fe_3O_4$. **(a)** Scanning electron microscopy (SEM) image of industrial $Fe_3O_4$ and **(d)**, transmission electron microscopy

(TEM) of nanoscale Fe$_3$O$_4$. *In-situ* X-ray diffraction of **(b)** industrial Fe$_3$O$_4$ and **(e)** nanoscale Fe$_3$O$_4$. Phase fraction of magnetite, wustite, and iron for **(c)** industrial Fe$_3$O$_4$ and **(f)** nanoscale Fe$_3$O$_4$ derived from *in-situ* X-ray diffraction results. The uncertainty for the analysis is less than 3%.

Iron reduction kinetics is known to exhibit strong dependence on the rates of adsorption and diffusion of reactive species. The initial reaction begins with H$_2$ surface adsorption and bond cleavage (*24*), but progresses via oxygen atom diffusion toward the surface along the kinetically favored directions (following Fick's laws). Studies have observed the FeO phase's metastability causes it to proceed via Fe$_2$O$_3$ to Fe$_3$O$_4$ to Fe at temperatures below 570 °C, while passing through FeO intermediates between Fe$_3$O$_4$ and Fe at higher temperatures. Our Fe$_3$O$_4$ nanoparticles had much faster reaction rates, with the majority of nanoscale Fe$_3$O$_4$ being reduced to FeO at 800 °C within 3 minutes (**Figure 1f**). For a mechanistic picture to compare with established literature, we explored this chemistry at lower temperatures, using synchrotron-based approaches.

**Low-temperature reduction of nanoscale Fe$_3$O$_4$.** The high X-ray flux and efficient X-ray detection at the Advanced Photon Source allowed us to resolve the low-temperature formation and consumption of FeO quite clearly, using the setup shown schematically in **Figure 2a** (*25*). Representative XRD traces from our experiment at 300 °C are plotted in **Figure 2b**, with markers at the bottom indicating the representative positions of the diffraction peaks characteristic of each phase. We unexpectedly observe the FeO phase forms preferentially in all reactions observed, even down to 300 °C. Many studies thus far have demonstrated that the FeO phase is thermodynamically unstable below 570 °C (*26*), though it forms kinetically from 450-570 °C, likely due to kinetic competition (*16*). Our observation of the preferred formation of FeO down to 300 °C indicates further kinetic stabilization in nanoparticles than has been observed or predicted previously (*27, 28*).

By comparing traces (**Figure 2c**) showing the formation of iron at 800 °C, 350 °C, and 300 °C, our results indicated three characteristic modes of these kinetics. At 800 °C, we observed rapid conversion to FeO, completed in 3 minutes. At intermediate temperatures (350 °C), we observed a gradual conversion to iron, following a roughly logarithmic function, growing towards an asymptotic population from the beginning of the reaction time. At the very lowest temperature, however, we observe a lag time in the formation of metallic iron lasting ~200 minutes, after which the iron forms following an exponential distribution.

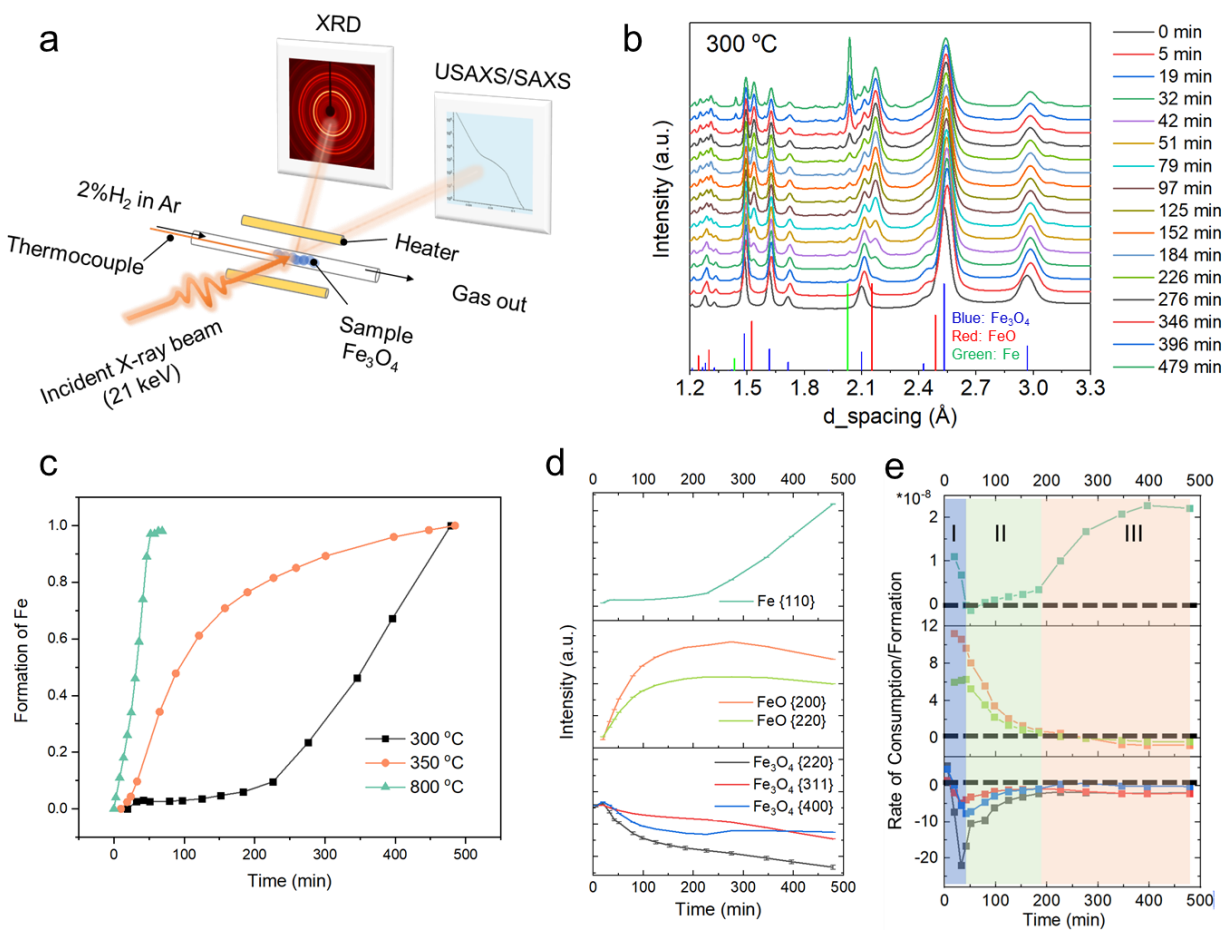

**Figure 2**. Overview of *in-situ* XRD of chemistry for Fe₃O₄ nanoparticles at 300° C. **(a)** A schematic of *in-situ* small angle X-ray scattering (SAXS), and X-ray diffraction (XRD) experimental configuration. **(b)** Representative time traces of *in-situ* XRD showing the nanoparticle chemistry of Fe₃O₄ at 300° C in 2% H₂ in Ar (g). **(c)** The plot of the rate of formation of iron at 300° C, 350° C, and 800° C. The uncertainty for the analysis is less than 3%. **(d)** The phases and facets of the anomalous 300° C chemistry are shown by the normalized integrated intensity under the relevant {hkl} diffraction peaks. **(e)** The plot of the first derivative of the traces from (c), indicating the net consumption and formation of each phase and crystal plane as a function of time.

To investigate the reason for the exponential growth at 300 °C, we study the full progression of the consumption and formation of Fe₃O₄, FeO, and Fe as shown in **Figure 2c**. The area under the curve for each peak in the XRD traces from **Figure 2b** is related to information about the volume abundance of crystals oriented in that way. For the fully isotropic powder in this experiment, temporal changes to the relative area under each peak indicate crystallographic changes to the volumes present for those directions. We include a full description of our normalization procedures and interpretation of the synchrotron X-ray diffraction analysis in **Supplementary Note 1**. By taking the derivative of the area under each peak with respect to the time of the integrated intensities, we measured the rate of consumption/formation of each phase, as shown in **Figure 2d**. The plots in **Figure 2d** demonstrate rates of consumption for points with negative values, and rates of formation with positive values, all with magnitudes that scale with rate. In the case of magnetite and wustite, the reaction rates differed for different diffraction peaks, indicating that the crystallographic orientation is intrinsically linked to the mechanism and associated kinetics.

As highlighted by the blue, green, and orange boxes in **Fig. 2d**, we observed 3 distinct behaviors of the phase transition dynamics that we interpret to be the stages of the chemistry. For each stage, we observe changes in the rate of conversion and the facet dependence of the chemistry. Starting from the perspective of phase conversion, in Stage I (the first 19 minutes), the $Fe_3O_4$ is consumed while both the FeO and Fe form *simultaneously*. In Stage II, we observe the consumption of $Fe_3O_4$ and net formation of FeO, with almost no change in the consumption/formation of Fe. Fe begins forming in Stage III, at $t = 200$ minutes with $Fe_3O_4$ and FeO both being consumed at similar rates. Beyond the phase considerations, the reaction rates on each crystallographic plane differ for each stage of the chemistry. In Stage I, the consumption of $Fe_3O_4$ is fastest along the {220} plane (gray), while the FeO formation is fastest along {200} (orange). By contrast, in Stage II, the net rates of formation for the {200} and {220} (green) facets of FeO converge, while the $Fe_3O_4$ consumption along {220}, {311} (red), and {400} (blue) initially separate before coalescing by the start of Stage III. In Stage III, the rates for each crystallographic facet differ minimally for each oxide phase.

Facet-dependent rates of chemistry like we observe in Stage I/II have been observed in other systems to occur based on the surface energy, which drives a change in the chemical driving force (Gibbs free energy) that governs the kinetics of the reaction *(29, 30)*. Magnetite's inverse spinel structure is often modeled as an FCC lattice of oxygen sites; this indicates surface energy trends of $\gamma_{111} < \gamma_{100} < \gamma_{110}$ *(31)*. We project the three planes measured in **Fig. 2d** onto the characteristic {111}, {110}, and {100} surface planes in **Supplementary Table 1** to verify that the {311},

{220}, and {400} planes measured from diffraction are consistent with those planes' kinetics, respectively. As such, the trend that the rate is fastest along {220}, intermediate along {400}, and slowest along {311} is consistent with the trend of the highest surface energy driving the fastest kinetics. In wustite's rock-salt structure, the surface energy trends as $\gamma_{100} < \gamma_{110}$, which indicates the {110} facet has the strongest driving force for chemistry. We observe the wustite phase forms fastest along the {200} facet and slows back to zero more rapidly, which is also consistent with our findings.

We interpret that the initial formation of *both* FeO and Fe in Stage I is caused by rapid chemistry at the nanoparticle surface, since there are sufficient active surface oxygens that reacts with $H_2$ directly without mass transport. The strong dependence of the rate on facet-energy in Stage II indicates the sustained mechanism of $Fe_3O_4$ to FeO conversion is influenced strongly by the surface energies. The role of surfaces is complex in HyDR reactions, which are known to be sensitive to many parameters (*16*), making the native reduction mechanisms nontrivial to establish. Despite variation in the literature, the $Fe_3O_4$ to FeO transition has precedent in being modeled as a surface-dominated reaction (*26*), while the FeO to Fe transition is usually described as diffusion-limited. To demonstrate this mechanism, we use a simplified mechanism based on a multi-step shrinking-core model to describe the competing reduction mechanisms (**Supplementary Note 2**). The reaction kinetics are known to exhibit competition between the surface reaction and the diffusion of atoms through the bulk product layer (*32*). Based on the model, we estimated the diffusivity of oxygen atoms across the product Fe layer and the reaction rate of surface oxygen atoms with $H_2$ (**Supplementary Note 2**). The high ratio between the surface reaction rate and atomic diffusivity (Thiele number $\Phi^2 \sim 10^{17}$) suggests a significant diffusion-driven behavior. We thus predict a hindered formation behavior of Fe that consists of what we observed during Stage II.

As the reduction process transitions from Stage II to III, the Fe phase starts to form, while the reaction rate of Fe and $Fe_3O_4$ phases remain nearly constant. This behavior indicates that the diffusion-driven mechanism of FeO-Fe reaction does not hold anymore. On the other hand, early in Stage III, the net reaction rate of the FeO phase crosses over from net formation to net consumption. We note that while we were not able to directly measure facet dependence in the iron phase, the significant narrowing of the peak widths in XRD indicates a significant increase in the effective particle sizes based on Scherrer broadening. The size increase and loss of facet dependence in the rate of consumption for both oxide phases seemed reminiscent of the whiskering mechanisms observed elsewhere (*33*). This led us to study how the dynamics in Stage III coupled to nano- to micro-scale structural changes.

**Microscopic view of Stage III Magnetite Reduction.** To directly map the changes in size as the nanoparticles react, we measured simultaneous small-angle X-ray scattering (SAXS) with our XRD from **Figure 2**. SAXS measures an integrated view of the average representative length scales of ensemble populations of particles and voids in samples with characteristic size populations (e.g. nanoparticles). Our experiments measured alternating XRD and SAXS patterns to resolve the average particle size across the measured 6-mm$^3$ volume inside the reactor capillary. The SAXS progression in **Figure 3a** plots *in-situ* scattering traces collected during the **Fig. 2b-d** reaction, from the start (purple) to the 485 minute end (red) of the experiment, measured based on the Q-values of $\frac{2\pi}{d}$ to resolve the populations for each $d$. The bump in the trace shown signified by the black arrow indicates a Guinier distribution that resolves a population of 10-nm diameter particles. As shown in **Figure 3a**, the emergence of features at lower Q values with increasing reaction time (shown by the red arrow) indicates the formation of larger features as the reaction progresses (~ 50 nm by the end of Stage III). Our SAXS results indicate a size progression proceeds simultaneously with the reduction chemistry shown in **Figure 2**.

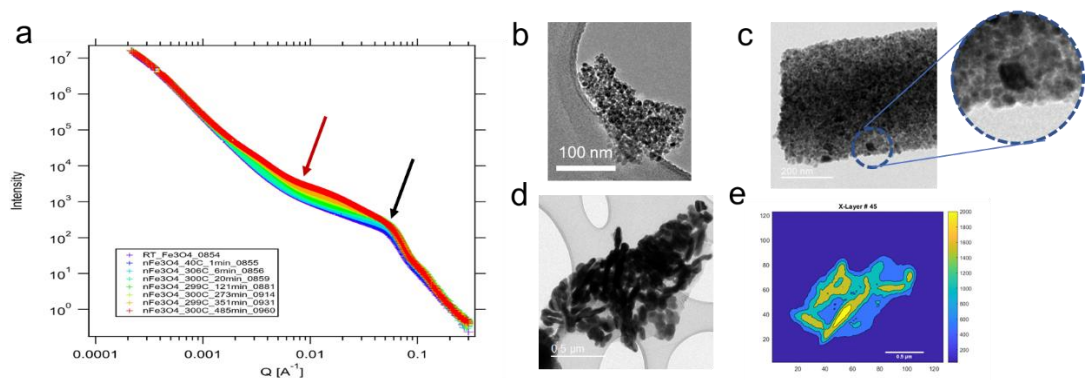

**Figure 3**. Size increases measured from reactors. **(a)** *In-situ* SAXS of reduction reaction shown in Fig. 3b for 300 °C reaction in 2% H$_2$. Transmission electron microscopy images of **(b)** the initial Fe3O4 nanoparticles, **(c)** the intermediate material after 2.5 hrs and, **(d)** the final oblong product material after reacting for 8-hrs. **(e)** 2D extracted slice showing a contour plot from the X-ray

ptychography 3D images after 8 hours. All images were collected from reactions at 400 °C in 2% $H_2$.

To guide our SAXS interpretation, we performed *ex-situ* imaging of samples extracted from our reaction vessel at different stages of the reaction (0-hrs, 2.5-hrs, 8-hrs; all collected at 400 °C). **Fig. 3b-d** shows how our TEM images revealed a structuring mechanism at the particle scale. We observed that the initial randomly-packed nanoparticles (**Figure 3b**) appear to self-assemble at the start of the reaction into rod-like structures (**Figure 3d**), that then coalesced into larger elongated grains of 100-350 nm length and 20-50 nm width (**Figure 3c**). For the statistical analysis, see **Supplementary Note 3**. As analogous control experiments in Ar (g) form isotropically random particle geometries (**Supplementary Figure 9**). We interpret that the reactive $H_2$ gas and associated phase transformations cause the formation of preferentially elongated structures in **Fig. 3**.

We confirm that the nano-rod structures (**Figure 3**) observed by TEM were also arose from increased phase density using X-ray ptychographic tomography, as shown for a representative 2D slice in **Figure 3e** (extracted from a 3D reconstruction). The analogous oblong structures over a 1-µm length scale corroborates the preference for these linear structures, and the higher electron density confirms the dominantly iron structure of the nano-rods.

We present the full 3D X-ray ptychographic tomography structures in **Figure 4** of the electron density integrated per voxel (Fig. 5a-c: 36-$nm^3$/voxel, Fig 5d: 18-$nm^3$/voxel). For direct comparison between these figures, we include a histogram of the binned populations of each electron density value, binned over all voxels measured for each of the three reaction stages (**Figure 4a**). The peaks of each histogram show the most prominent electron density values for each stage of the reaction, which is determined by the phase of the particles and the packing density of the particles. For our starting material, we identify a peak (black) (~600 e/$nm^3$) that indicates a mean packing density of 36 $Fe_3O_4$ particles per voxel, indicating the loose packing of initial particles (i.e. ~55% fewer than the maximum number of particles arranged in a close-packed structure). While the lower valued peak indicates the iron oxide particles fuse to eliminate void space, the higher value can only correspond to the higher electron-density Fe phase beginning to form in dilute and small domains as the particles fuse. After $t = 8$ hours, the peak (blue) of ~1400 e/$nm^3$ indicates the conversion to the iron product, with a continued increase in packing density in the new Fe phase. We note that the broadness of the peak likely also indicates a significantly wider variance on packing densities in iron, as well as partial re-oxidation and/or incomplete chemistry. See **Supplementary Note 4** for a full description of our ptychography analysis. The iron nucleus formation observed in the histograms in **Fig. 4a** are consistent with the TEM images in **Fig. 3**, as illustrated by the inlay in **Fig. 3d**. From these findings, we interpret that the self-assembled particles begin the fusion/conversion process at the end of Stage II, as seen by the anomalously dark regions.

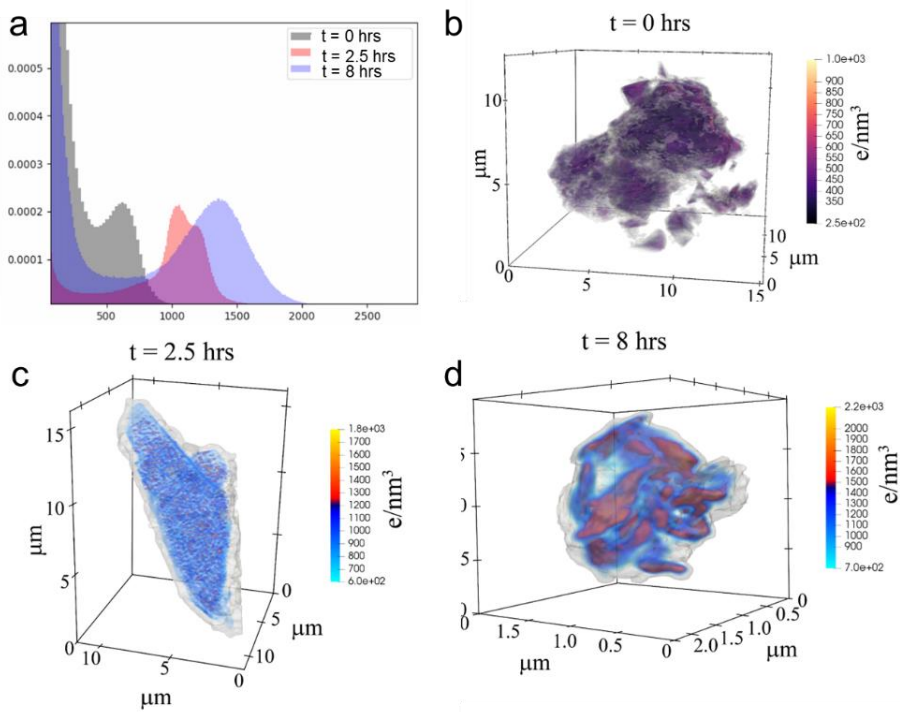

**Figure 4.** X-ray ptychography of the sintering process during nano-$Fe_3O_4$ reduction in 2% $H_2$ in Ar at 400 °C. **(a)** Histograms of integrated voxel electron densities plotted in **black, red, and blue** for samples extracted after t=0-hrs, t=2.5-hrs, and t=8-hrs. X-ray Ptychography results of **(b)** initial $Fe_3O_4$ nanoparticles; edge artifacts were removed for clarity **(c)** intermediate sample and **(d)** nanoworm structures in final iron product.

The full 3D spatial maps of the packing densities of representative structures are shown for the starting material (**Figure 4b**), the halted reaction at $t = 2.5$ hours (**Figure 4c**) and the sample collected after $t = 8$ hours (**Figure 4d**). We note that **Figure 4d** is a higher resolution scan than the one given in the histogram to show the finer features representative of the rod-like structures (which were integrated at larger scales). The 3D structure for the blue trace from the histogram is shown in **Supplementary Figure 12**. Images in **Figure 4** show isosurfaces displaying the 3D contours of constant electron density for the *highest* electron density regions in the material at the same representative stages in reaction as in **Figure 3**. We display the increase in electron density by the color/transparency scaling shown in each figure. The initial samples show no preferred packing arrangements, as the domains of densest packing are random in size and orientation (also observed in **Fig. 3b**). In comparison, the intermediate timescale reveals a number of domains for the high electron-density domains, each with roughly spherical shapes that are representative of the nuclei in **Fig. 3a**. The final image from the reaction shows rod-like linear structures that are analogous to the TEM structures in **Figure 4e** and representative of the lengthscales observed by SAXS. The electron-density distributions in **Figure 4a** connect this sintering mechanism directly to the phase transition to form iron.

From our imaging results, we observe that our initially random assemblies of nanoparticles undergo a transition during Stage I-II to self-assemble into the elongated structures in **Fig. 3c**. A few small regions on the edges of the rod-like assemblies appear to be abnormally dense, as shown by the abnormally dark regions **(Fig. 3c)**. By the end of Stage III, the particles are entirely the dark region, indicating a completion of that process. As such, by comparing our imaging results with the Stages shown in **Figure 2**'s diffraction results, we conclude that the sintering mechanisms observed in this work seed during the *second* stage then grow during Stage III, i.e. the wustite to iron transition. Through this connection and the strong dependence of Stages I-II on the crystallography, we conclude that the chemistry is strongly influenced by crystallography, and the phases formed are intrinsically linked to the sintering mechanisms. This implies the surface and reaction interface energies may both connect to crystallographic anisotropy.

**Discussion & Conclusions**

Previous thermodynamics models have illustrated that surface energies of iron oxides can change the relative energy stabilities of the different iron oxide phases in <100 nm particles. In nanoparticles, the high surface to volume ratio amplifies this effect, however, the wustite phase has been predicted to be destabilized by that effect. Our observations illustrate that despite thermodynamics, kinetic effects dominantly drive the $Fe_3O_4$ to FeO transition *below* the bulk stability of the FeO phase at 570 °C. This implies that the surface-stabilizing kinetics previously proposed extends to an even lower temperature during the $Fe_3O_4$ reduction in hydrogen - either due to stabilization at phase boundaries or other mechanisms.

The tubular structures our images resolve is characteristic of the "whiskering effect" often observed in HyDR of natural iron ore samples(*33*). Similarly, the facet dependence of the chemistry we observe is frequently observed in other nanoparticle systems. As we show in **Figure 1a,** nanoparticles are present in natural iron ore fines used in industry, indicating that the unexpectedly formed FeO phase we observe by *in-situ* XRD at 300 °C is relevant to anomalous cold spots generated in reactors today, though it is currently overlooked. While the nanoparticles are small, their preference for self-assembly and sintering indicates possible nucleation sites and/or barriers to the reduction that has not been explored in reactor-scale models. Nucleation sites are the highest-energy portion of most kinetic pathways, and is thus essential to describe macroscopic nucleation and growth processes to connect the atomic-scale chemistry to the sintering known to degrade reactor performance.

Fundamental studies like ours are essential to inform the science to optimize the chemistry and mass transport essential to optimize processes that rely on iron oxide redox chemistry. Many of these process designs, however, overlook the science underlying the sintering effects at unprecedentedly low temperatures. By bypassing the magnetite to hematite transition known to form pores and vacancies that propagate inherited defects and alter kinetics, our results reveal complex structuring in ribbon shapes formed by particles representative of the smallest lengthscales in iron ore fines. Our findings provide key insights into the kinetics and thermodynamics of this complex mechanochemistry giving a perspective into scales that are necessary to model "unconventional" feedstocks (i.e. bypassing the formation of pellets). Fundamental studies like ours are essential to inform the science to optimize the chemistry and mass transport essential to design and optimize new clean ironmaking processes.

## Methods

**$Fe_3O_4$ nanoparticle synthesis.** The $Fe_3O_4$ nanocrystals were synthesized by a high-temperature solution-phase hydrolysis reaction (*34*). NaOH stock solution (2.5 mmol/L) was prepared by dissolving 50 mmol NaOH in 20 mL diethylene glycol (DEG), followed by heating at 120 °C for 1 h under nitrogen and cooling down and keeping at 70 °C. FeCl$_3$ stock solution (0.4 mmol/L) was prepared by dissolving 8 mmol FeCl$_3$ in 20 mL DEG, followed by heating at 80 °C for 1 h and cooling down to room temperature. In a typical procedure, 288 mg polyacrylic acid (PAA), 5 mL of FeCl$_3$ stock solution, and 10 mL DEG were mixed and heated to 220 °C in a nitrogen atmosphere for 30 min under vigorous magnetic stirring, forming a transparent light-yellow solution. 4.5 mL of NaOH stock solution was injected rapidly into the above hot mixture, inducing the temperature drop to 210 °C and the reaction solution turned black immediately. The resulting mixture was further heated for 10 min. Then another 5 mL of FeCl$_3$ stock solution was injected into the above reaction. Upon the reaction temperature raised up to 220 °C, 3 mL of NaOH stock solution was added. After 15 min, Fe$_3$O$_4$ nanocrystals with size of around 10 nm were obtained. The final products were cooled down to room temperature and washed by repeated actions of precipitation with ethanol and subsequent redispersion in deionized (DI) water several times, and finally redispersed in DI water. Fe$_3$O$_4$ nanoparticles in DI water were fully dried at 60 °C under vacuum and used as powder.

**Synchrotron X-ray measurements.** *In-situ* ultra-small angle X-ray scattering (USAXS), small angle X-ray scattering (SAXS), and X-ray diffraction (XRD) are carried out at beamline 9-ID-C from the Advanced Photon Source. Gas of 2% H$_2$ in Ar was flowed through the sample, which was packed into quartz capillaries with an inner diameter of 0.9 mm. Continuous gas flow was monitored using a bubbler, and the nanoparticles were held in place with glass wool. Temperature is controlled with a temperature accuracy of ±5 °C. X-ray ptychographic tomography was performed on the Velociprobe (*37*) located at 2-ID-D beamline at the APS. Pure magnetite nanoparticles, intermediate and final samples at the three reduction stages ($t$ = 0, 2.5, 8 hours, at 400 °C with 2% H$_2$ gas in Ar) were prepared on Si$_3$N$_4$ windows with a membrane thickness of 200 nm. An 8.34-keV monochromatic X-ray beam was focused by a Fresnel zone plate with an outermost zone width of 50 nm, with an illumination size of about 1.5 μm on the sample plane. Ptychographic projection scan was carried out in a raster fly-scan pattern with a horizontal step size of 150 nm and a vertical step size of 300 nm. Far-field coherent diffraction patterns were recorded by a Dectris[1***] Eiger X 500K detector (2.335-m downstream of the sample) at a

---
[1***] Certain commercial products or company names are identified here to describe our study adequately. Such identification is not intended to imply recommendation or endorsement by the National Institute of Standards and Technology, nor is it intended to imply that the products or names identified are necessarily the best available for the purpose.

continuous frame rate of 100 Hz. For 3D scan of each sample, 315 ptychographic projections were acquired at angles ranging from -79 ° to 78° with an angular step of 0.5°.

**Characterization.** Scanning electron microscopy images were collected using an FEI Magellan 400 XHR SEM with FEG source. Transmission electron microscopy images were taken at FEI Tecnai G2 F20 X-TWIN Transmission Electron Microscope with a field-emission gun (FEG). The Empyrean X-ray Diffractometer from PANalytical was used for lab-based X-ray diffraction data collection.

**Acknowledgments**

This work was funded in part by the Stanford Doer School Sustainability Accelerator award, under contract 270636. This work was supported by the U.S. Department of Energy (DOE), Office of Basic Energy Sciences, Division of Materials Sciences and Engineering (Contract No. DE-AC02-76SF00515). Part of this work was performed at the Stanford Nano Shared Facilities (SNSF), supported by the National Science Foundation under award ECCS-2026822. This research used resources of the Advanced Photon Source, a U.S. Department of Energy (DOE) Office of Science user facility operated for the DOE Office of Science by Argonne National Laboratory under Contract No. DE-AC02-06CH11357. The authors are also thankful to Yue Jiang and Xiaolin Zheng, (Department of Mechanical Engineering, Stanford University) for their help with control experiments.

**Additional information**

Supplementary information is available in the online version of the paper. Correspondence and requests for materials should be addressed to L.D.M.

**Data Availability**

The data that supports this study are included in the article and/or supplementary information.

**Completing financial interests**

The authors declare no competing interests.

**Author contributions**

# Supplemental Information for: Correlating Chemical Reaction and Mass Transport in Hydrogen-based Direct Reduction of Iron Oxide

Xueli Zheng[1,5], Subhechchha Paul[1], Lauren Moghimi[1], Yifan Wang[1], Rafael A. Vilá[1], Fan Zhang[2], Xin Gao[1], Junjing Deng[3], Yi Jiang[3], Xin Xiao[1], Chaolumen Wu[4], Louisa C. Greenburg[1], Yufei Yang[1], Yi Cui[1], Arturas Vailionis[1], Ivan Kuzmenko[3], Jan llavsky[3], Yadong Yin[4], Yi Cui[1,5], Leora Dresselhaus-Marais[1,6]*

[1]Department of Materials Science and Engineering, Stanford University, Stanford, California 94305, USA.

[2]Materials Measurement Science Division, National Institute of Standards and Technology, Gaithersburg, MD 20899, USA.

[3]X-ray Science Division, Advanced Photon Source, Argonne National Laboratory, Argonne, Illinois 60439, USA.

[4]Department of Chemistry, University of California, Riverside, CA 92521, USA.

[5]Stanford Institute for Materials and Energy Sciences, SLAC National Accelerator Laboratory, Menlo Park, California, USA.

[6]Department of Photon Science, SLAC National Accelerator Laboratory, 2575 Sand Hill Rd, Menlo Park, 94025, CA, USA.

*Correspondence and requests for materials should be addressed to L.D.M (leoradm@stanford.edu)

**This PDF contains the following information.**
**Supplementary Notes 1-4**
**Supplementary Figures 1-13**
**Supplementary Tables 1-2**

**Supplementary Note 1. Interpretation of Diffraction Experiments**

The analysis of our X-ray diffraction data collected at the synchrotron is explained in this note. We first list the givens that are important to understand in our interpretation of the data. Firstly, our experiments were conducted in powder-diffraction geometry, meaning they measure only a small part of the Debye Scherrer ring. The rings we observed in all cases were completely homogeneous in intensity along the width of the camera, and when measured on an area detector in a lab source were consistent with no crystal texture over the course of the reaction. As such, we interpret our data in terms of 2θ without consideration for $\eta$. Additionally, all peaks in our starting materials were significantly broader than the conventional bulk diffraction signal, which is consistent with Scherrer broadening for particles of < 100 nm diameter sizes. Identification of the atomic structure of phases with XRD is typically done with Rietveld analysis, which uses statistical methods to fit the atomic positions based on a comparison between the simulated and observed diffraction patterns. It is well established in the literature that Rietveld analysis is contingent upon a deterministic scattering model, which often is not possible with nanoparticle samples. While we did attempt Rietveld analysis for our interpretation of the data included in **Figure 3** of the main text, the results of the algorithm in all cases gave results that were nonphysical - either due to poor fitting or inappropriate atomic positions.

As such, we use fitting algorithms for peaks of each phase that are clearly isolated to perform our interpretation of the data. As surface strains and crystal truncation effects are known to broaden and change the diffracted intensities in nanoparticle diffraction, we interpret the volume of scattering planes in our crystal based on the integrated area under the fitted peaks that we measured for each plane. We divide the measured area under each magnetite peak by that collected on the nominal pure intensity collected at $t = 0$ to account for the structure factor and {$hkl$} multiplicity that are known to alter the scaling of each peak. The resulting values we report are thus consistent with the relative volumes of each scattered plane of the crystal; as the relative intensities of the peaks change, it shows that the number of planes for each {$hkl$} value changes with respect to each other, indicating that the number of planes with that symmetry is lessened asymmetrically. When translated to the rate of consumption/formation of a given phase, this indicates the facet-dependent chemistry we report in our manuscript.

**Supplementary Note 2. Diffusion-surface reaction competition in early stage magnetite reduction.** To understand the anomalous early-stage magnetite reduction behavior in our experiments, we propose a simple mechanism based on the multi-step shrinking core model *(6)*. Given the 10-nm size of our synthesized $Fe_3O_4$ nanoparticles, it is reasonable to assume that the gas-solid reaction undergoes a multi-step shrinking core mechanism, as shown in **Supplementary Figure 6a-d**. Stage I begins with the consumption of pure $Fe_3O_4$ and formation of both FeO and Fe. The product FeO and Fe are nucleated at multiple locations on the surface, while there are still abundant active oxygen sites of magnetite on the outer surface, as shown in Supplementary Figure 6e. At this stage, the rate is governed by the surface chemical reactions, as reported in Guo et al *(7)*:

$H_2 + O^{2-} + 2Fe^{3+} \rightarrow H_2O + 2Fe^{2+}$

$H_2 + O^{2-} + Fe^{2+} \rightarrow H_2O + Fe$

This overall reaction rate as a function of temperature $T$ can be estimated by the rate-limiting step $H_2(g) + 2O^* \rightarrow 2^*OH$ with the Arrhenius relation,

$r^* = A \exp(-E_b/k_BT)$

where $k_B$ is the Boltzmann constant, the prefactor A = 277/s for 2% $H_2$ gas, and the energy barrier $E_b$ = 0.4352 eV *(6)*.

At the end of Stage I, the nucleus of FeO and Fe join together, forming a thin layer of α-Fe at the outer surface (**Supplementary Figure 6b**). At this point, there is no active oxygen on the surface, and the process is primarily driven by diffusion: Oxygen atoms O* formed at the FeO-Fe surface require to diffuse through the product layer (α-Fe lattice) to reach the surface, as shown in **Supplementary Figure 6f**. This mechanism can be validated by estimating the dimensionless Thiele number $\Phi^2$ *(8)*, defined as the ratio between the surface reaction rate and diffusion,

$\Phi^2 = 4\pi r_0^2 \, \Gamma_{O^*} \, r^*/D_{O^*}$

where $\Gamma_{O^*}$ = 1.135×10$^{19}$ (m$^{-2}$) is the fraction of active oxygen sites per unit area (estimated based on the oxygen atoms on the $Fe_3O_4$ {001} surface), $r_0$ = 5nm is the radius of the particle, and $D_{O^*}$ is the diffusion coefficient of O* atom in the Fe lattice, which can be estimated by

$D_{O^*} = D_0 \exp(-Q/k_BT)$

where $D_0 = 3.78 \times 10^{-7}$ m$^2$/s and $Q = 0.96$ eV *(9)*.

At the low temperature (300 °C) experiment where Stage I and II present, the surface reaction rate and the diffusion coefficient are estimated as $r^* = 0.0413$/s and $D_{O^*} = 1.3601 \times 10^{-15}$ m$^2$/s. Thus, the Thiele number is calculated as $\Phi^2 \sim 10^{17}$, demonstrating a strong diffusion-driven mechanism at the end of Stage I and during Stage II.

During Stage II, we observed slow formation of Fe, but fast consumption of Fe$_3$O$_4$ and formation of FeO, as shown in **Figure 3d**. Here, we propose a diffusion-driven mechanism to understand this anomalous behavior as shown in **Supplementary Figure 6f**. the formation of Fe now undergoes Fe$^{2+}$ + O$^{2-}$ → Fe + O$^*_{(lat)}$ at the Fe-FeO interface. Here, Accumulation of O$^*$ near the Fe-FeO interface hinders the formation of Fe formation rate, shown as Stage II in **Figure 3d**. There is a concentration gradient of O$^*$ atoms within the thin layer of Fe due to the low diffusivity of $D_{O^*}$. The concentration of O$^*$ is low near the outer surface and high near the Fe-FeO interface, driving an outward diffusion of O$^*$ atoms. Although the Fe formation rate is hindered, FeO continues to form from Fe$_3$O$_4$ at a relatively high rate (**Figure 3b**). In the FeO and Fe$_3$O$_4$ phases, the O$^{2-}$ ions consist of a closed-pack face-centered cubic (fcc) structure. The diffusion coefficient of O$^{2-}$ in the wustite phase is extremely low (D = $1 \times 10^{-30}$ m$^2$/s at 400 °C) compared even to O$^*$ diffusion in Fe product layer. Therefore, the electron transport through the wustite phase is achieved by iron (Fe$^{2+}$) vacancy diffusion in the wustite phase from the FeO-Fe$_3$O$_4$ interface to the Fe-FeO interface *(10)*. During actual reation process, wüstite Fe$_{1-x}$O is usually presented to be a non-stoichiometry phase with Fe$^{2+}$ vacancies. These free vacancies can move around easily in the wüstite phase *(11)*. The reaction mechanism can then be written as O$^{2-}$ + v$_{Fe}^{2+}$ → O$^*$ at the Fe-FeO interface. At the FeO-Fe$_3$O$_4$ interface, the Fe$^{2+}$ vacancy in the wustite phase is equivalent to the octahedral site v$_{oct}$ in the Fe$_3$O$_4$ phase, and the electrons carried from the outside is then transported to Fe$^{3+}$, i.e., Fe$^{3+}$ + $e^-$ → Fe$^{2+}$.

The abovementioned diffusion-driven mechanism explains the initial "delay" of the iron formation during Stage I and II. As the reaction transitions from Stage II to III, our ex-situ TEM and Ptychography observations suggest that there is a significant nanoparticle agglomeration. We propose that the microstructure change to the nanoparticles, thus break the diffusion-driven mechanism explained above. At intermediate (350 °C) and high (800 °C) temperatures, the agglomeration nucleates faster (*1*). As a result, Stage I and II disappears at these temperatures as shown in Figure 3c.

**Supplementary Note 3. Transmission Electron Microscopy (TEM) analysis.** TEM scans of starting and final materials are analyzed using image segmentation in MATLAB. For the TEM scan of the starting material, we select a 2D region of interest representative of the volume element of Ptychography data (i.e. 36 nm in our case) as shown in Fig. S8.

The histogram in Fig. S9a shows the distance, *d*, between all pairs of points identified in the TEM image from Figure S8, scanned over 39 points collected from a single image. The bimodal distribution shown in Fig. S10a has one peak at $d < 10$-nm and one peaked at $d > 10$-nm. Since our 5-nm radius particles could only have had a minimum separation distance of $d = 10$-nm in the same 2D plane, we therefore interpret that the in-plane separations in our measurements were represented by the distribution peaked at $d > 10$-nm. Moreover the peak at d>10nm behaves like an outlier and is neglected in further calculations. In order to effectively fit the data, we truncated the distribution, *f(d)*, at $d = 10$-nm, then fitted it to a Gaussian distribution as shown in Fig. S9b. The fitted data with a mean of 5 nm and a variance of 6 nm, shows a preference for a separation of 17 nm in between particles of the starting material. We incorporate this separation value in our electron density calculations to compute the maximum voxel electron density (MVED) of the starting material, to get an idea about the initial packing as observed via TEM. Comparing it to the distribution obtained from Ptychography data, we note that the MVED value from TEM lies within 2 standard deviations of the Ptychography images.

For the TEM scan of the final material (shown in Fig. S10a ), we located features of different sizes on the image and plotted their major and minor axis lengths as a histogram in Fig. 10b. The histograms represent the wide distribution in size of the elongated structures.

**Supplementary Note 4. X-ray ptychography analysis**. In this study we combined x-ray ptychography with tomography to measure the volumetric refractive index of our samples at 3 stages during reduction (t=0, 2.5, 8 hours, at 400 °C  with 2% hydrogen gas in argon), and then further calculated the 3D electron distribution within the samples. After obtaining 3D ptychographic data, 2D ptychographic complex-valued projections were firstly reconstructed by a GPU-based code ith a real-space resolution unit up to18 nm. Because the phase information is more sensitive in hard X-rays, only phase parts were used for tomographic reconstruction. The obtained phase projections were then aligned to a common rotation axis and tomography reconstructions were carried out by simultaneous iterative reconstruction technique (SIRT) using ASTRA toolbox. The tomographic results gave 3D distribution of the real part of the refractive index of the sample, $\delta(\vec{r})$, which was used to calculate the 3D electron distribution by

$$n_e(\vec{r}) = \frac{2\pi\delta(\vec{r})}{r_e\lambda^2}, \qquad [1]$$

where $r_e$ is the classical electron radius and $\lambda$ is x-ray wavelength.

While the electrons per cubic nanometer are more general, the voxelization at the 36 nm$^3$ scale is important to interpretation, as it indicates the coarseness over which the packing density of the material may be evaluated at. In other words, sub-36-nm$^3$ particles cannot be distinguished as objects with shape, but as effective packing densities based on modeling assumptions about the occupied and unoccupied volumes in each voxel as recorded from ptychographic imaging for the three different stages of the reaction. Fig. S11 demonstrates how different size particles might occupy a voxel of a particular size and how it affects the packing density and void spaces.

The peaks of the distribution for each histogram in Figure 5a show the most prominent electron density values that are described by each voxel, which is related to the packing density and phase of the material present. While the reaction is known to progress via Fe$_3$O$_4$ →FeO → Fe, the literature reports that the FeO phase is unstable at room temperature. As such, our interpretation assumes the presence of only magnetite and Fe phases, with packing particle densities that are indicative of the sintering mechanisms during the reaction. In order to understand the packing density in the voxels, it is necessary to estimate the maximum number of nanoparticles (MNP) that can be packed in each voxel. We assume a hexagonal close packing (hcp) structure and spherical nanoparticles of known radii. Pure magnetite particles of 5 nm radii are used as starting material. The radius of pure iron is computed to be ~4 nm considering volume contraction of reactants as the reaction progresses. These assumptions allow us to calculate the volume of hcp unit cells for each phase, the ratio of which with the voxel volume returns the maximum number of hcp unit cells that can be packed in a voxel of known size. Since there are six nanoparticles per hcp unit cell, one can easily interpret the MNP which occupies the voxel volume. The number of electrons present in a single nanoparticle of a particular phase is extracted from the knowledge of the unit cell structure of pure magnetite and pure iron. The product of the MNP and the number of electrons in a single nanoparticle (for each phase) yields the maximum number of electrons (MNe) present in the voxel. The ratio of (MNe) to the voxel volume returns the maximum voxel electron density (MVED). The MVED for each phase is then compared with the peak values of the histogram plot to estimate approximately the actual number of nanoparticles of each phase that occupy the voxels. Table S2 lists the values used for the computation.

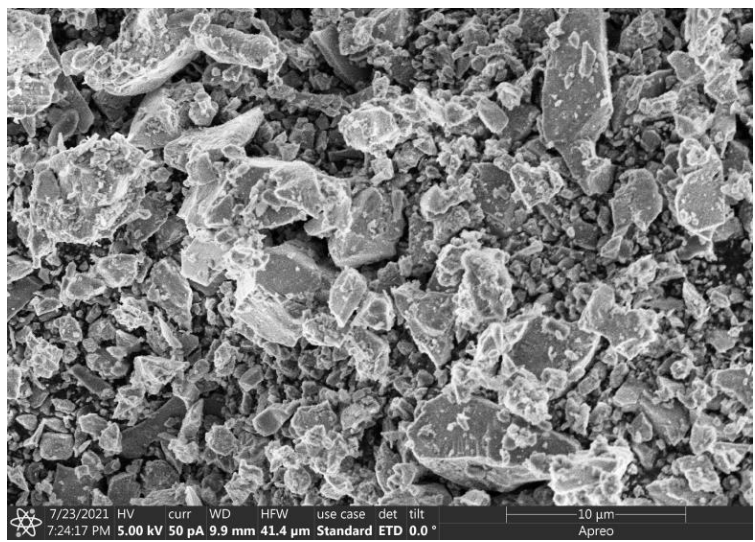

**Supplementary Figure 1.** Scanning electron microscopy image of industrial magnetite.

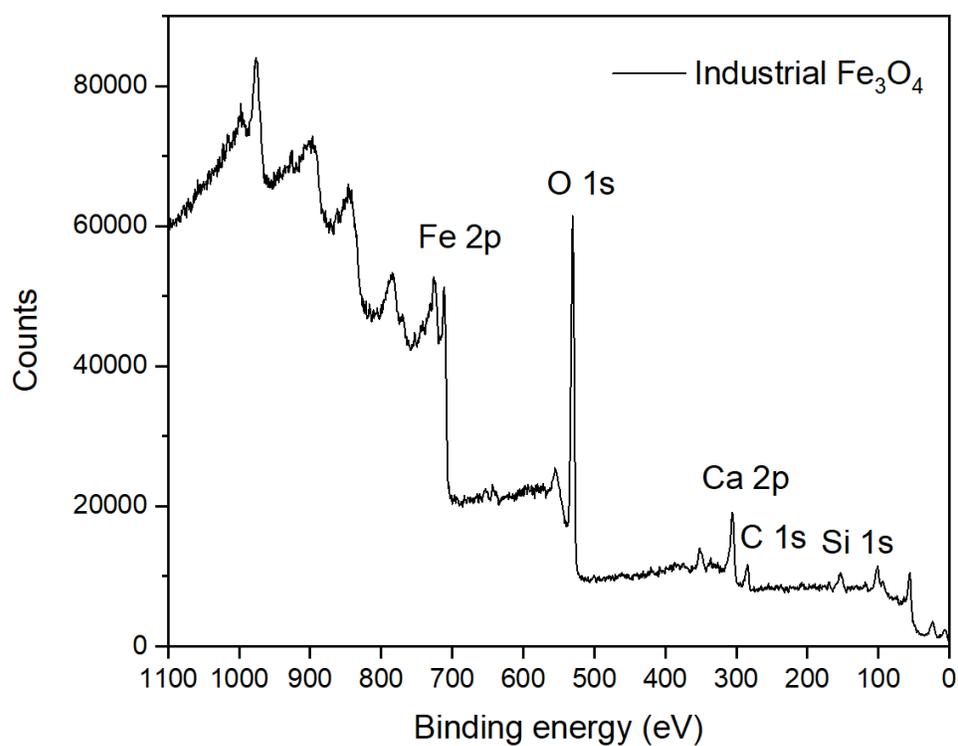

**Supplementary Figure 2**. X-ray photoemission spectroscopy of industrial Fe3O4.

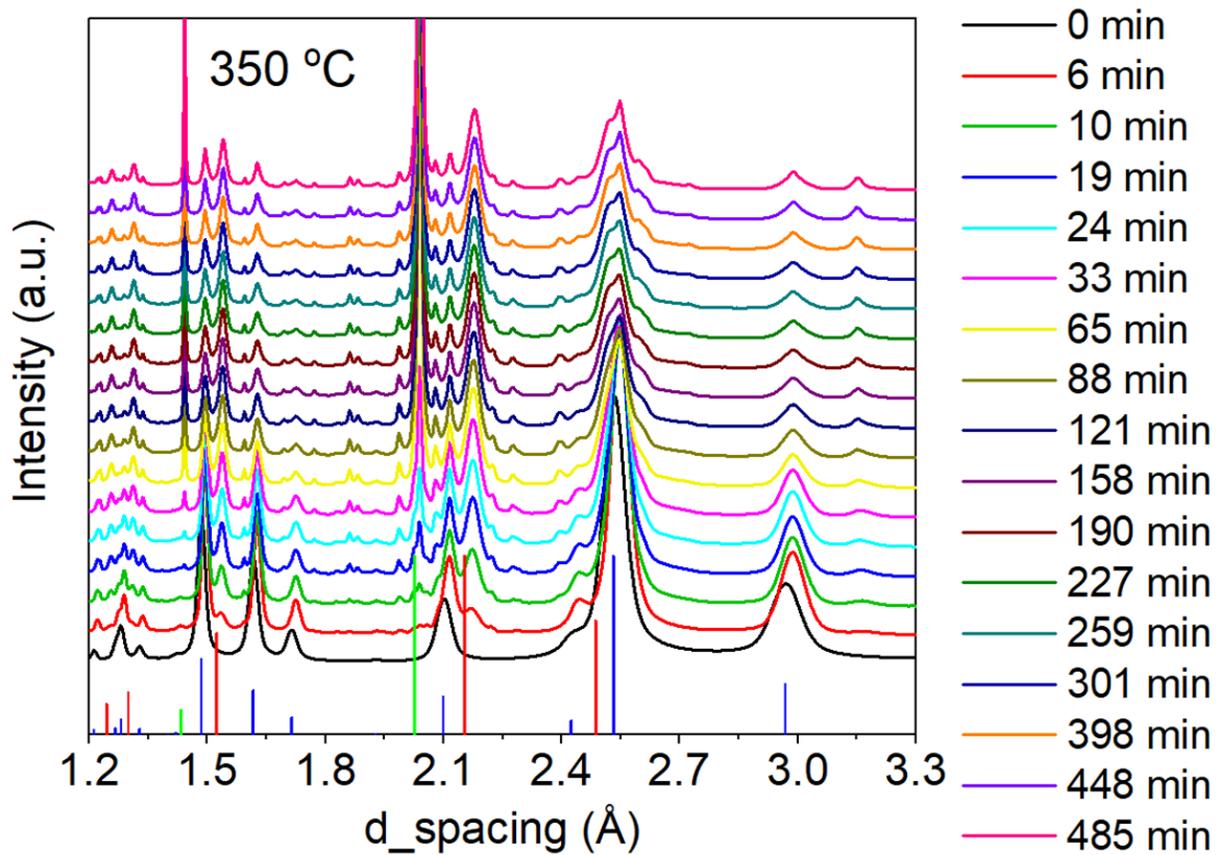

**Supplementary Figure 3.** In situ XRD traces from 350C.

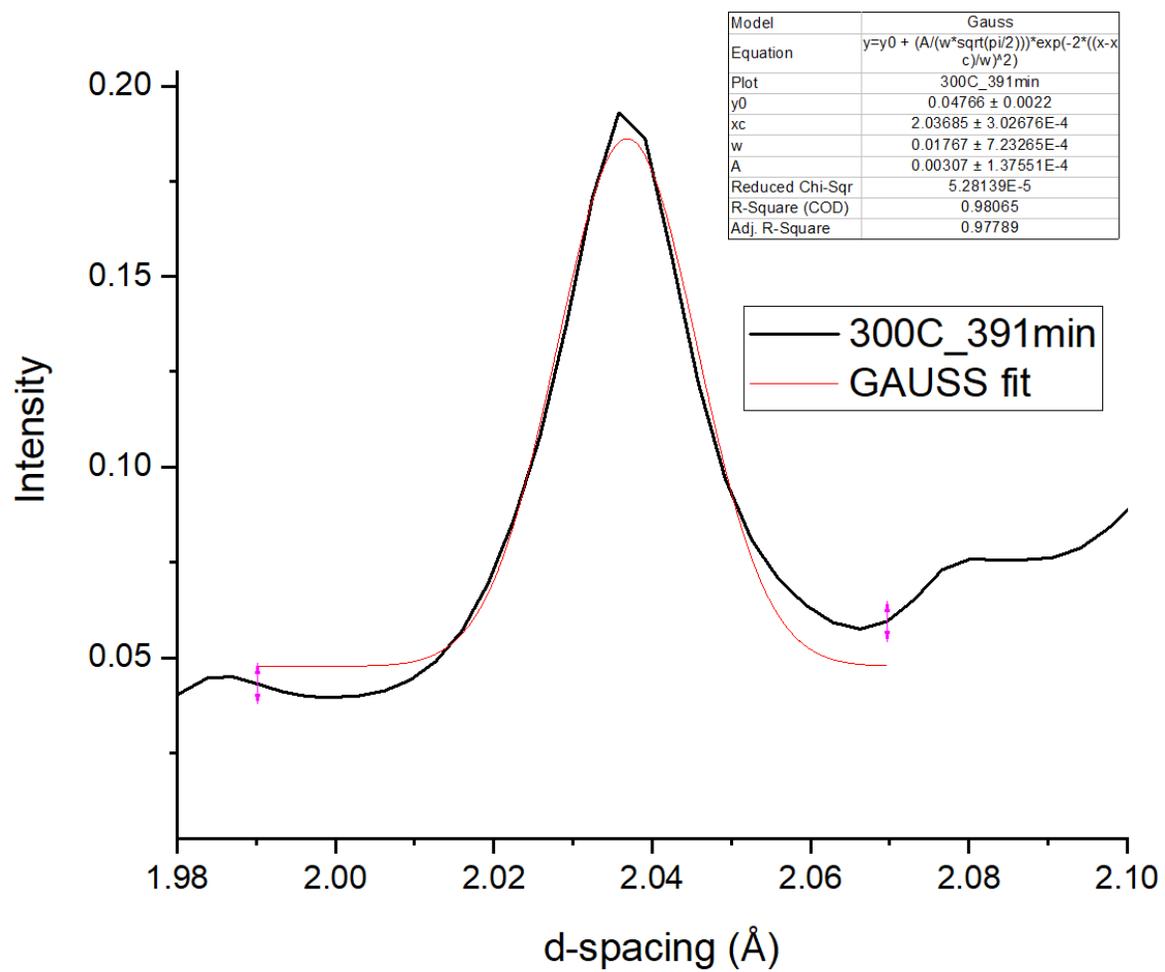

**Supplementary Figure 4.** Gaussian fitting of Fe peak at tested 300 C.

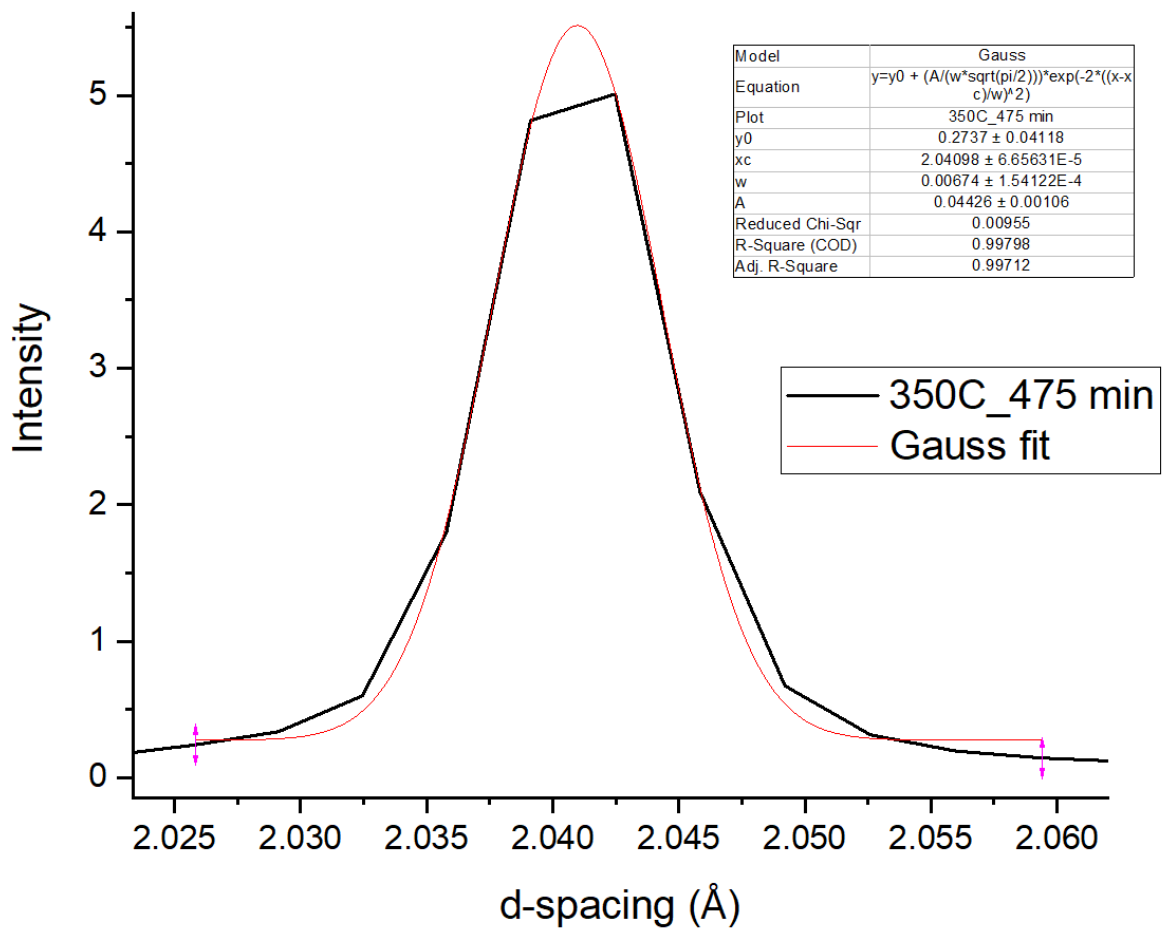

**Supplementary Figure 5**. Gaussian fitting of Fe peak at tested 350 C.

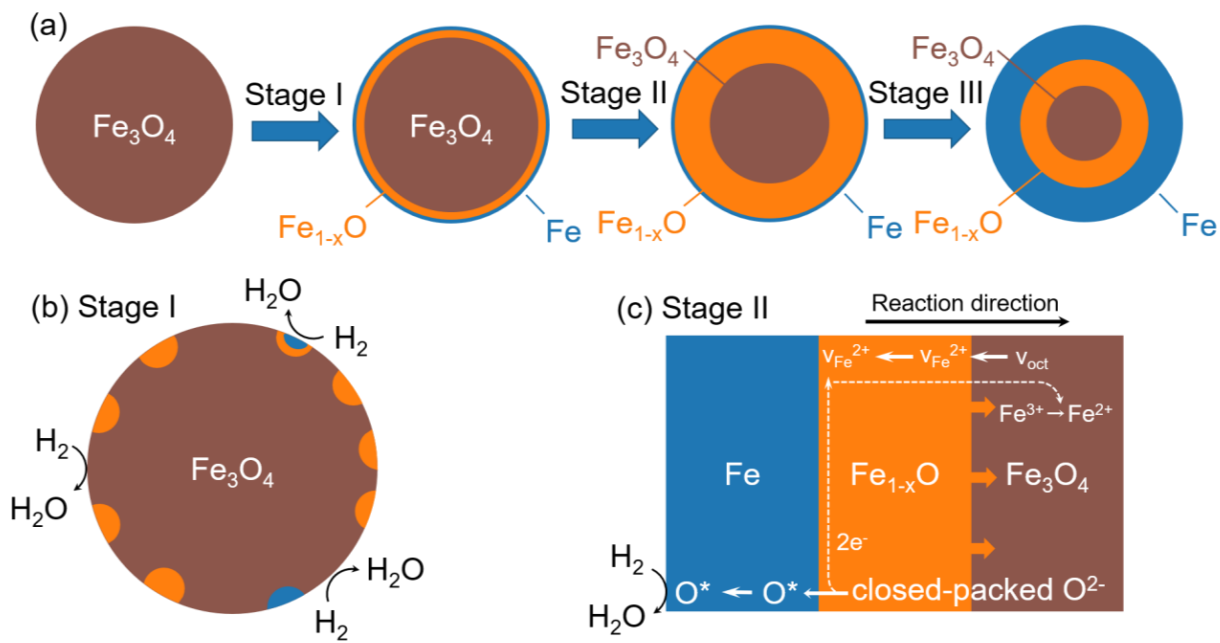

**Supplementary Figure 6.** (a) Multi-step shrinking core model describing the three stages of magnetite reduction. (b) Multiple nucleation clusters of wustite and iron on the surface of magnetite nanoparticles. (c) Reduction pathway during reduction Stage II, where only the reaction front between wustite and magnetite is moving (thick orange arrow). White arrows indicate the diffusion (Oxygen atom O* and iron vacancy $v_{Fe}^{2+}$) through the lattice, and the dotted line indicates the transition path of electrons.

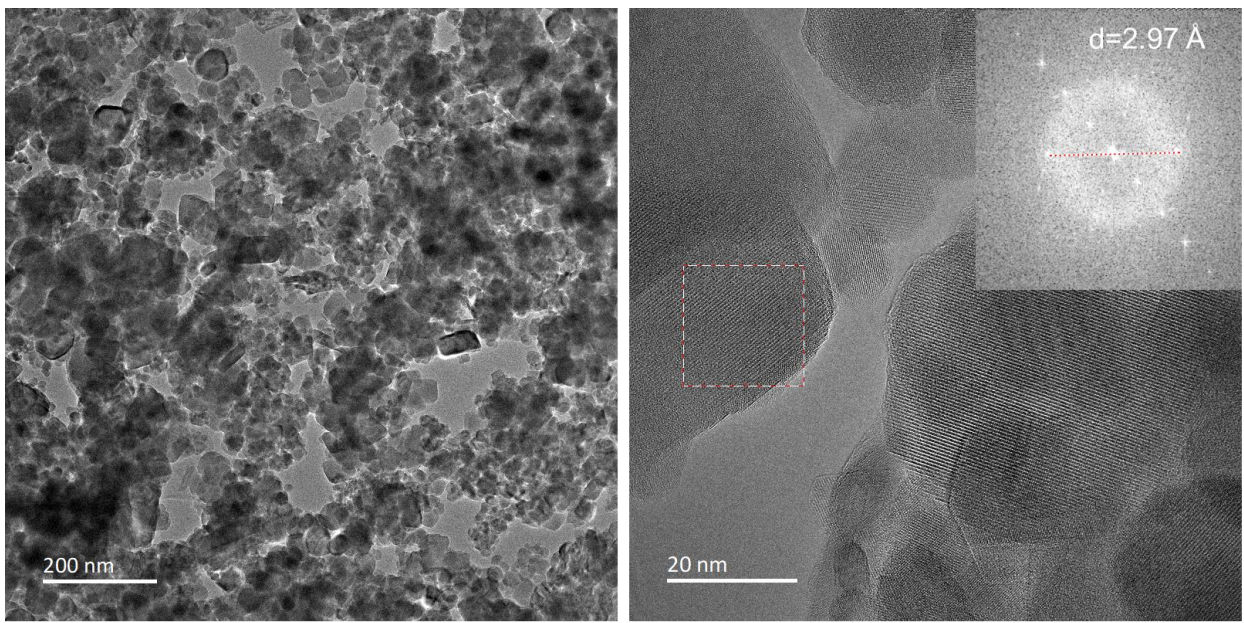

**Supplementary Figure 7**. *ex-situ* TEM images of control experiment, i.e. after heating Fe3O4 for 8 hrs at 400 oC under an inert atmosphere of argon gas. The left image shows the overall structure of the particles, while the right image shows a high-resolution TEM image with atomic resolution. The inset in the right image shows the Fourier transform of the region selected by the dashed line box, which indicates the existence of {220} of Fe3O4 with d-spacing of 2.97 Å.

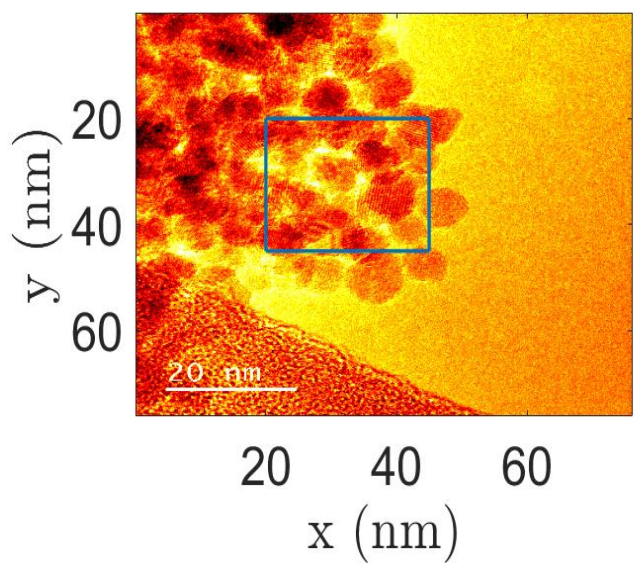

**Supplementary Figure 8.** TEM scan of starting material, i.e. 10 nm pure magnetite particles. The box encloses a region of interest of size 25nmx25nm.

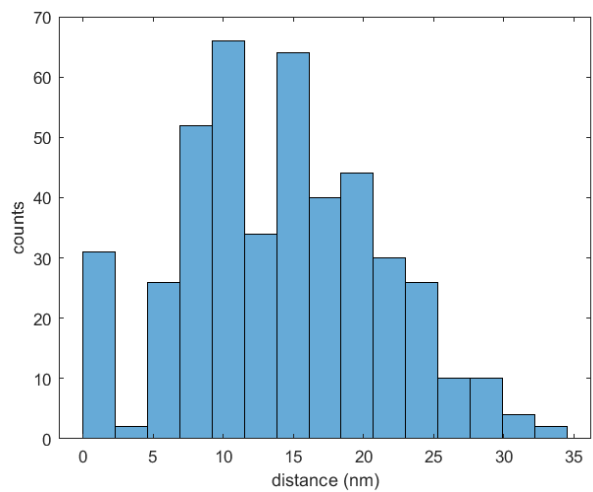

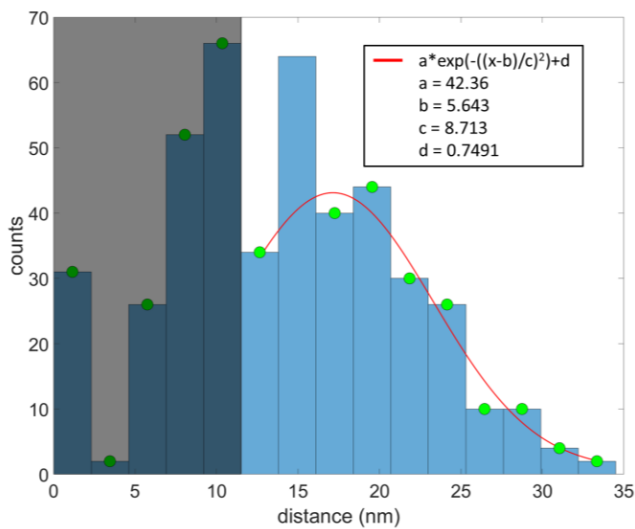

**Supplementary Figure 9.** (a) Histogram of starting material (TEM scan) with 15 bins on 39 particle centroid points collected for the starting-material particles shown in Fig. S3. b) Gaussian fitted data along with the initial histogram for the starting material. The shaded region shows the excluded points for $d < 10$-nm, while the green markers indicate the points used for the fit. We excluded the bin at $d = 15$-nm as a statistical outlier to enable the fit to most effectively describe the data.

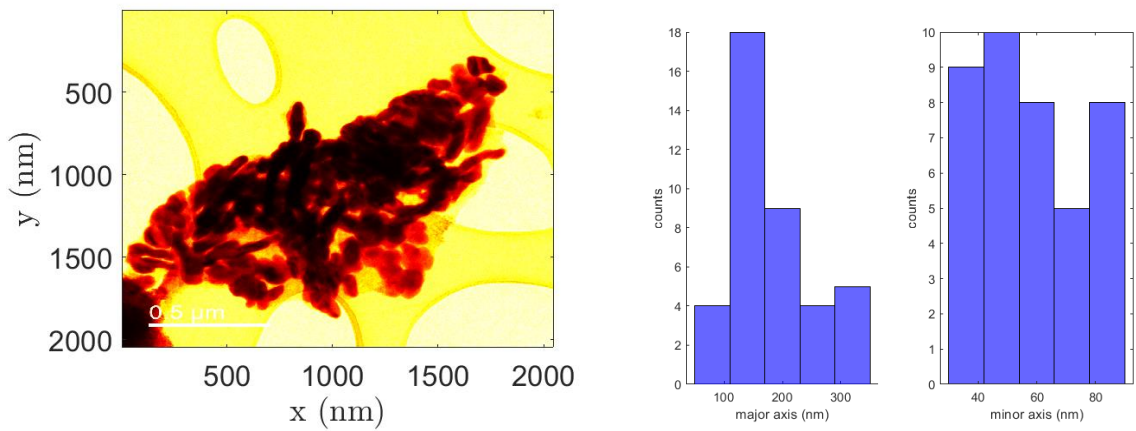

**Supplementary Figure 10. a)** ex-situ TEM scan of pure magnetite nanoparticles reacted with 2% $H_2$ in the Ar atmosphere for 8 hours showing elongated structures. b) Histogram of TEM scans.

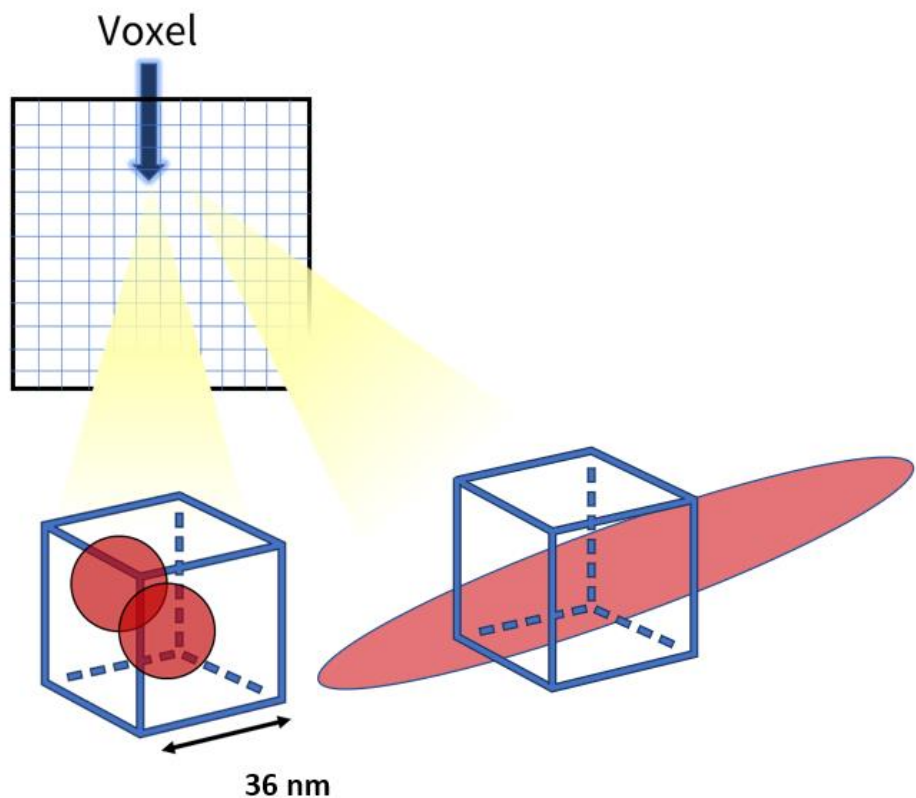

**Supplementary Figure 11.** Representation of a single voxel enclosing nano-particles (left) and containing elongated features (right) formed at the final stage of the reaction.

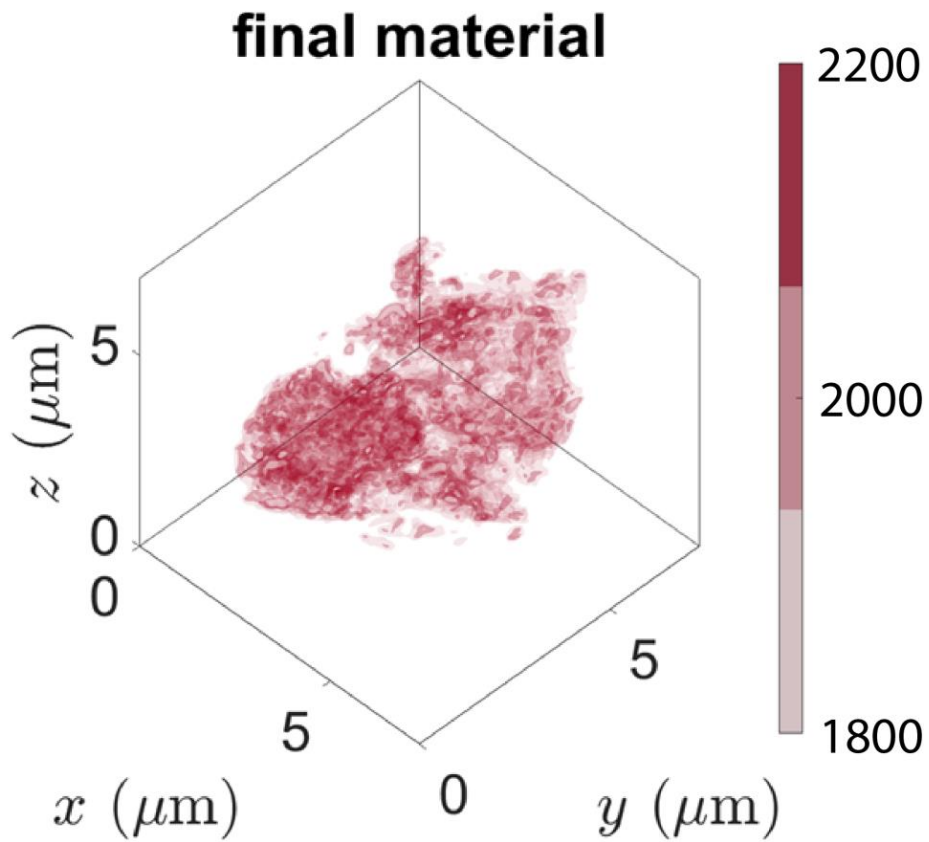

**Supplementary Figure 12.** Ptychography image of the 36-nm$^3$ trace shown in the histogram in Fig. 5a of the main paper. We note that while the 3D structure is similar to the starting material in this view, the nano-elongated structures observed in the structures shown in the main text were difficult to resolve clearly with the larger integrations.

**Supplementary Table 1.** Projections of the observed {400}, {220}, and {311} planes onto the known facets to describe surface energy. We note that the highest valued projection of the diffraction plane onto the principal *hkl* vectors describes the rate of chemistry that is most strongly represented by that diffraction analysis.

|  |  | **Diffraction Plane** | | |
|---|---|---|---|---|
|  | *hkl* | {400} | {311} | {220} |
| **Surface Plane** | {111} | 0.57735 | 0.927173 | 0.816497 |
|  | {110} | 0.707107 | 0.648886 | 1 |
|  | {100} | 1 | 0.229416 | 0.707107 |

**Supplementary Table 2.** Calculations of the predicted electron densities for the starting phase of magnetite and final phase of iron, assuming nanoparticles.

|  | **Magnetite** | **Iron** |
|---|---|---|
| Unit Cell Structure | Inverse Spinel | Body Centered Cubic |
| Nanoparticle Radius (nm) | 5 | 4 |
| Nanoparticle Separation Based on TEM Data (nm) | 17 | — |
| MNP/voxel | 66 | 128 |
| MVED (nm$^{-3}$) | 1175 | 1623 |
| MVED TEM (nm$^{-3}$) | 239 | — |
| Peak Values (nm$^{-3}$) | 600 | 1400 |
| Packing density: $\frac{Peak\ Values}{MVED} \times 100\%$ | 51 | 85 |